\documentclass[a4paper,11pt]{article}
\def\ll{\label}
\def\re{\ref}
\def\c{\cite}

\def\r1{(\ref{$1})}

\def\ba{\begin{array}{c}}

\def\ea{\end{array}}

\def\de{\delta}
\def\bet{\beta}
\def\ov{\over}
\def\ha{{1\over 2}}

\def\l{\left}
\def\l({\left(}
\def\r){\right)}
\def\r{\right}

\def\la{\lambda}
\def\al{\alpha}

 \def\be{\begin{equation}}
\def\bc{\begin{center}}
\def\ec{\end{center}}
\def\bit{\begin{itemize}}
\def\eit{\end{itemize}}
\def\ee{\end{equation}}
\def\ed{\end{document}}
\def\bea{\begin{eqnarray}}
\def\eea{\end{eqnarray}}
\def\efr{\end{flushright}}

\begin{document}
\title{Quantum integrability and 
Bethe ansatz solution  for interacting  matter-radiation systems}
%\vskip 1cm

\author{
Anjan Kundu \\  
  Saha Institute of Nuclear Physics,  
 Theory Group \\
 1/AF Bidhan Nagar, Calcutta 700 064, India.
\\  {email: anjan@tnp.saha.ernet.in} }
\maketitle
\begin{abstract} 
%---------------------------------------------------
A  unified   integrable  system, generating a new series of
interacting matter-radiation models with interatomic coupling and different
atomic frequencies, 
  is constructed and exactly  solved  through  algebraic Bethe ansatz. 
 Novel features in Rabi oscillation and vacuum Rabi splitting are shown on
the example of an integrable two-atom Buck-Sukumar model with resolution of
some important controversies
 in the Bethe ansatz solution  including its possible
 degeneracy  for such models.
%67-----------------------------

% interacting photon_atom (like 42.50-p (q-optics)

\end{abstract}

 The basic physics underlying a variety of  important 
phenomena in interacting  matter-radiation   (MR) systems, 
like those in quantum optics induced by
resonance interaction  between atom and a quantized  laser field, 
 in cavity QED
 %both in microwave and optical domain 
\c{rempe8790,raizen89}, 
in trapped ion
interacting with its center of mass motion
 irradiated by a laser beam \c{trap,vogel95} etc., seems to be nicely
 captured by  simple models like  Jaynes-Cummings (JC)  \c{jc},
Buck Sukumar (BS)  \c{bs} and some of their extensions  \c{jc1}.
Many theoretical predictions  based on these models, like
 vacuum Rabi splitting (VRS) \c{raizen89,VRS},
 Rabi oscillation  and its  quantum collapse and revival \c{rempe8790}  
etc. 
 have been verified in maser and laser experiments.
 However, for describing physical situations more accurately one has to look
  for further generalizations of the basic models, like q-deformed BS  
and JC  model \c{qbs,qjc}, trapped ion  (TI) with nonlinear coupling
\c{vogel95,ntrap}, 
 multi-atom models \c{raizen89,ntrap,natom} etc.
 Nevertheless, while the exact solutions for the  JC and the  BS models
together with
their simple multi-atom extensions
 are known \c{jcexact,jcbethe,bsbethe},  the
 same is no longer true
for most of the above generalizations.
Moreover, while in known multi-atomic MR models the atoms interact
only via the oscillator mode \c{jcbethe,bsbethe} with coinciding
 atomic frequencies (AF),  integrable models 
with explicit inter-atomic
 couplings  
 have not  been proposed.   Likewise, though q-deformation, 
which physically signifies  introduction of anisotropy together with
specific
nonlinearity into the system,
was considered for a few MR models \c{qbs,qjc}, their multi-atom and
integrable variants 
  are not  known.
 Therefore, it is indeed a challenge to find a  scheme
for generating integrable  MR models with 
the desired properties.

 To meet this challenge we construct  a general 
 integrable system based on the
ancestor Lax operator of \c{kunprl} and generate in a unified way a series
of integrable multi-atom MR models with explicit inter-atomic interactions
and nondegenerate AF.
This includes such new
 generalizations for JC, BS, TI, etc. models and discovers 
important integrable q-deformations like  qBS,  qJC, qTI etc. It
is worth noting that, our integrable TI model  exhibits full
 exponential nonlinearity without any approximation 
and multi-atom qBS and qJC models involves quantum group
 spin operators. 
Moreover, since our  construction
 is based on  a general
 Yang-Baxter  (YB) algebra, together with the  generation of 
various models at its different realizations, we 
  can solve
them exactly in a unified  way through  algebraic Bethe ansatz (BA).
 Our strategy of construction is to start with 
  a Lax operator by taking  it as a  combination
 $T(\la)=L^{s}(\la) \prod_j^{N_a}L_j^{S}(\la)$,
with  $L^{s}(\la)$ linked with the ancestor model of  \c{kunprl} 
 and  $N_a$-number of $L_j^{S}(\la)$ related to the spin model   \c{jcbethe}.
 By construction  
it must satisfy the YB equation 
 $R(\la-\mu)T(\la)\otimes T(\mu)= (I\otimes T(\mu))
(T(\la)\otimes I)R(\la-\mu)$, with  mutually commuting set of conserved
operators 
%$[C_a,C_b]=0$
  obtained from the expansion   $
\tau (\la)= tr T(\la)=\sum_a C_a \lambda ^a $ \c{aba}.
For standard  MR models, as we will see below,  the Lax
operators are  rational type linked with the simplest quantum
  $R$-matrix of 
$xxx$  spin chain \c{aba}, while  for  $q$-deformed  models they 
are  trigonometric type  related to the  $R$-matrix of $xxz$ chain 
 \c{xxz}. We
concentrate  first on standard MR models and recall that 
in the rational case the $2\times 2$ ancestor Lax operator may be given 
as
\be
L^{s}{(\la)} = \left( \begin{array}{c}
 {c_1^0} (\la + {s^3})+ {c_1^1}, \ \ \quad 
  s^-  \\
    \quad  
s^+ ,  \quad \ \ 
c_2^0 (\la - {s^3})- {c_2^1}
          \end{array}   \right), \ll{LK} \ee
with operators $ {\bf s}$ satisfying  a 
 quadratic algebra 
\be  [ s^+ , s^-] 
=  2m^+ s^3 +m^-,\ 
  ~ [s^3, s^\pm]  = \pm  s^\pm,\ \ [m^\pm,\cdot]=0. \ll{ralg}\ee 
The central elements   $m^\pm $ are 
expressed through
 arbitrary parameters appearing
 in (\re{LK}) as $m^+=c_1^0c_2^0,\ \  m^-= c_1^1c_2^0+c_1^0c_2^1$ and as  
 it is easy to see, their different  choice  reduces (\re{ralg})
 to  different algebras:
\bea
\mbox {i)  } su(u),  \ \mbox {at }  \ 
  m^+=1, m^-=0, \quad  
 \mbox {ii)  } su(1,1),  \ \mbox {at }  \ 
  m^+=-1, m^-=0,  \nonumber \\  
 \mbox {iii)  bosonic}, \ \mbox {at }  
  m^+=0, m^-=-1, \ \mbox {iv)  canonical},  \mbox {at }  
  m^+= m^-=0  \ll{algs}\eea 
and the corresponding limits yield  from (\re{LK})  the  respective Lax
operators. In case i), (\re{LK}) reduces simply to the spin Lax operator
\be
L^{S}_j{(\la)} = \left( \begin{array}{c}
 \la + {S_j^z}+ {c_{j}}, \ \ \quad 
  S_j^-  \\
    \quad  
S_j^+ ,  \quad \ \ 
\la - {S_j^z}+ {c_{j}}
          \end{array}   \right), \ll{LS} \ee
Our  Lax operator constructed as above would generate the set of 
all  commuting conserved 
operators, 
with  higher ones containing  increasingly  higher many-body
 interactions.   The simplest among them is
$ C_{N_a}= s^3+ \sum_j^{N_a}   S^z_j$, while   the next  in the set  $
 \al C_{N_a-1}$, may be  defined 
  as the Hamiltonian
 of our unified MR system: 
\bea
H_{{ MR}}&=& H_d+ H_{ S s}+H_{S S}, \nonumber \\ 
H_d&=&\omega_f s^3+ \sum_j^{N_a}  { \omega_a}_j S^z_j ,\nonumber \\  
 H_{Ss }&=& \al \sum_j^{N_a}\left (
 s^+ S^-_j+ s^- S_j^+ +(c^0_1+c^0_2) s^3S^z_j \right ) 
,\nonumber \\
 H_{SS}&=& \al \sum_{i < j} \left((c^0_1+c^0_2) S^z_iS^z_j+ c^0_1 S^-_iS^+_j+ 
c^0_2 S^+_iS^-_j \right)
\ll{imrh}\eea
Here
 $H_{Ss }$
 describes  matter-radiation, while
 $H_{SS } $,  
matter-matter interactions.
 ${\bf S}_j, j=1,2,\ldots, N_a$ stand
 for an array of $N_a$ atoms, each with $2s+1$ levels
 and satisfy the $su(2)$ algebra. ${\bf s}$ on the other hand 
signifies  a radiation or a 
vibration mode and  satisfies more
 general algebra
(\re {ralg}). In
 (\re{imrh}) the  radiation frequency $\omega_f $ and the  atomic 
 frequencies $\omega_{aj}, j=1,2, \ldots, N_a$
 are defined through  inhomogeneous 
parameters of the Lax
operator as 
\be \omega_f=\sum_j w_j, \ \ w_j=\al (c^0_1-c^0_2) c_{j}
, \ \ \  \omega_{aj}= \omega_f-w_j +\al (c^1_1+c^1_2)\ll{frec}\ee
 
Remarkably, 
the   general model  (\re{imrh})  reduces to a
new series of integrable multi-atom BS, JC and  TI models 
 in a unified way at the  limits ii), iii) and iv) of (\re{algs}). 
For example,   case ii) with the choice   
\be c_1^0=- c_2^0=1,  c_1^1=c_2^1 \equiv c,  \ll{cbs}\ee
  yields from  (\re{imrh}) the model
\be
H_{BS}= \omega_f s^3+ \sum_j^{N_a} \left( { \omega_a}_j S^z_j+
 \al (s^+ S^-_j+ s^- S_j^+)\right)  
+   \al \sum_{i < j}^{N_a} (  S^-_iS^+_j- S^+_iS^-_j),
 \ll{nbsh}\ee
 which with 
a bosonic 
realization of $su(1,1)$:  $ \ 
 s^+=\sqrt N b^\dag, 
s^-= b \sqrt N , s^3=N+\ha \ $ and the spin-$s$ operator  
${\vec S}=\ha \sum_k^{2s}{\vec \sigma}_k$,
would represent a new {\it integrable multi-atom  
 BS model} with inter-atomic interactions and different 
atomic frequencies. 
Note that at $N_a=1$, when 
 matter-matter interactions vanish and all  AF
 coincide,  (\re{nbsh})
  recovers
   the known    model  \c{bsbethe}.
%by representing  ${\vec S}_1=\ha \sum_k^s{\vec \sigma}_k$. 
    However we 
  solve below exactly through BA the more general   case with
nonvanishing  interatomic couplings and 
all different AF: ${ \omega_a}_j $,
as defined in (\re{frec}).

Similarly, a 
 new {\it  integrable multi-atom JC model}
 with matter-matter coupling is obtained   from the same (\re{imrh}) 
  under reduction
iii), consistent with 
$c_1^0=\al, c_2^0=0, c_1^1 \equiv 
c, c_2^ 1=-\al^{-1}  $ and  bosonic
realization 
$s^-=b, s^+=b^\dag, s^3= b^\dag b . $
 We do not
present here  explicit form of this  easily derivable Hamiltonian, 
which yields
the known  model \c{jcbethe} only at $N_a=1$,
  when interatomic couplings vanish and all AF become degenerate.

We can generate an {\it integrable TI model}
 with interatomic interactions, again from the same MR model 
 (\re{imrh}) at reduction iv),
by fixing the parameter values as $c_1^0=-1, 
 c_1^1 \equiv  c, c_2^ 0= c_2^1=0  $ and considering  consistent realization 
through canonical variables as $s^\pm=e^{\mp i x}, \ \ s^3=p+ x$. We
present here only its $N_a=1$ form
by a suitable combination with the other conserved quantity $C_1$: 
\be
H_{TI}= (\omega_a-  \omega_f)S^z+{S^z}^2+  
\al (e^{-ix} S^++e^{ix }S^-)+ H_{x p},
 \ll{tih}\ee
with $H_{x p}= \ha (p^2+ x^2)+ xp, \ {\vec  S} =\ha \sum_k{ \vec \sigma }_k,
 $ which   is  a   new integrable multi-atom TI
model with full exponential nonlinearity 
 without  approximation.
% which is linked interestingly with the 
% construction of the  well known Toda chain \c{kunprl}.

For constructing     
 {\it integrable q-deformed MR models} the strategy would be the same; only
  one has to start  now
from the  trigonometric type ancestor Lax operator 
involving  q-deformed  operators
and  associated with  $xxz$
$R$-matrix, the explicit form of which is given in \c{kunprl}.   
For simplicity we present
here only  $N_a=1$ case with the Hamiltonian 
  \bea
H_{qMR}&=& H_d+  (s_q^+ S^-_q+ s^-_q S_q^+)\sin \al ,  \nonumber \\ 
H_d&=&-ic_0 \cos (\al X) +c \sin (\al X), \  X=(s_q^3-S_q^z+\omega),
 \ll{qimrh}\eea
which represent  a new class of {  MR models} with ${\bf S}_q$ belonging to
the   quantum group
$U_q(su(2))$ and   ${\bf s}_q$ to a more general quantum  algebra
\c{kunprl}.
It is important to note, that ${\bf s}_q$  
 can  yield a variety of
   q-deformed  operators,  inducing  
 (\re{qimrh}) to  generate  a number of  
 physically relevant q-deformed
integrable MR models. 

For example, an {\it integrable  q-deformed BS model}
may be  constructed   from (\re{qimrh})  at $c_0=0 $, 
 by realizing ${\bf s}_q$ through q-oscillator:
$s_q^+=\sqrt {[N]_q} b^\dag_q, \ s_q^-=b_q\sqrt{ [N]_q}, \ s_q^3=N+\ha  $,
and  quantum spin operator  ${\bf S}_q$ 
 by using its  co-product  \c{xxz} : 
$S^\pm_q=\sum_j^{s} q^{-\sum_{k<j}\sigma^z_k}
\sigma^\pm_jq^{\sum_{l>j}\sigma^z_l}, \ S^z =\sum_j^{s}\sigma^z_j$.   
Note that at  $s=1$, we get an integrable
  version of an earlier  model \c{qbs}.

Similarly   the same 
general model (\re{qimrh}) with   choice $c_0=i, c=1$ and 
 realization $s_q^+= b^\dag_q, s_q^-=b_q, s^3=N $
yield a new {\it integrable q-deformation of the JC model}, while
under  reduction  $c_0=i, c=0$ and the same  
 realization through canonical operators as for the 
 TI model,   it generates an {\it integrable q-deformation of the TI
model}.
By taking higher $N_a$ values multi-atom integrable variants 
 of all the above
q-deformed matter-radiation models can be constructed.
% related to the  relativistic Toda chain \c{kunprl}.

 We emphasize that  all 
MR models presented here, similar to their unified construction, allow
 their exact  BA   solutions  also in a unified and
   almost   model-independent way.
In BA formalism the diagonal entries
 $\tau (\la)=T_{11}(\la)
+T_{22}(\la)$ 
 produce all   conserved operators, while the
 off-diagonal elements $  T_{21}(\la)\equiv B(\la)$ and $ 
  T_{12}(\la) \equiv C(\la)$ 
   act like
   { creation} and { annihilation} operators of 
 pseudoparticles  with the M-particle  state defined as 
 $|M>_B=B(\la_1) \cdots B(\la_M)|0>$ and the pseudovacuum 
   $|0>$  through
$ C(\la)|0>=0$.
The basic idea of algebraic BA \c{aba} 
is to find the eigenvalue solution:
$\tau(\la)|M>_B=\Lambda (\la, \{ \la_a\})|M>_B$, for which  
 diagonal elements $ T_{ii}(\la), i=1,2$ 
 are   pushed through the string of $B(\la_a) $'s toward  $|0>$,
 using the commutation relations obtainable  from the 
 YB equation.  Considering further the actions
$T_{11}(\la)|0>=\alpha (\la)|0>, T_{22}(\la)|0>=\beta (\la)|0> $, 
one   arrives finally  at the eigenvalue expression 
\be \Lambda (\la,\{ \la_a\}) =
 \al(\la)\prod_{a=1}^M f(\la-\la_a)
+
\bet(\la)\prod_{a=1}^M f(\la_a-\la),
\ll{Lambda}\ee
where $f(\la)$ is defined through the elements of the 
 $R$-matrix as $
{\la+\al \ov \la} $, for the rational and
  as  ${\sin (\la+\al) \ov \sin \la} $ for the
 trigonometric case.
Expanding $ \Lambda (\la,\{ \la_a\}) $ in powers of $\la$ we  obtain
the eigenvalues for all conserved operators including the Hamiltonian, where
the rapidity  parameters $ \{ \la_a\} $ involved can be determined  from the 
Bethe equations 
\be 
{ \al(\la_a) \ov \bet(\la_a)}=
 \prod_{b \not =a}{ f(\la_b-\la_a) \ov
 f(\la_a-\la_b)}
, ~~ a=1,2, \ldots , M,
\ll{be}\ee
which follow  in turn from the requirement of $|M>_B$ to be an eigenvector.
Returning to our models we find that, 
the major parts in  key algebraic BA relations 
(\re{Lambda}) and (\re{be}),  described
  by   $R$-matrix elements $f(\la)$, depend 
 actually on the class to which the models belong, rather than  on an 
individual  model.
 Therefore, for all standard MR systems including 
BS, JC and TI models, $f(\la)$ is given by its same rational form,
 while for all q-deformed models like qBS, qJC, qTI etc. 
by  its trigonometric expression.
The only   model-dependent parts  in these equations, 
expressed through $\al(\la)$ and $ \bet(\la) $    are
 determined  from 
 our   general Lax operator construction, which  
for the rational class   using (\re{LK}) and (\re{LS}) is 
 obtained as 
\bea \al(\la)&=&(c^0_1(\la+r)+c^1_1)\prod_j^{N_a}(\la-s+c_{j}),\nonumber \\
  \bet(\la)&=&(c^0_1(\la-r)-c^1_2)\prod_j^{N_a}(\la+s+c_{j}), \ll{albet}\eea
where  $r=<0|s^3|0>$ depends  on  particular realization 
of (\re{ralg}) and 
$s=-<0|S^z|0>$ denotes the atomic spin.
(\re {albet})  yields easily 
  the  needed forms for  BS, JC and TI  models,
 at the  corresponding choices
 of the parameters  like  (\re{cbs}), as  we have noted above.
Similarly, for  q-deformed models   quantum 
extension of (\re{albet}) together with the trigonometric form for
$f(\la)$ have to be considered.
For the solution of   TI and qIT models however 
 one has to adopt a bit
different approach close  to that of  the Toda chain  \c{faba},
 since pseudovacuum is difficult to
determine for such models.

For deriving  physical consequences from  our constructions, 
we consider integrable  
   two-level  
   multi-atom  BS model
  with inter-atomic couplings, by taking spin-$\ha$ operator
${\vec S}_j=\ha {\vec \sigma} _j$ in   (\re{nbsh}). 
 Using the full strength of the BA 
method exact  solutions for this multi-atom  model with arbitrary
 $N_a$ and  different $\omega_a$'s can be
  derived 
  from the same BA relations   (\re{Lambda})-- 
 (\re{albet})  for all excitations, by 
 just tuning the parameters involved to their required reduction (\re{cbs}).
We demonstrate some  novel features in Rabi oscillation and VRS
 by using the BA  solutions in the $N_a=2$ atom case of (\re{nbsh}).
The first excited
 energy spectrum    $E_1=\omega_f+2 \la_1$  linked with 
 the   cubic  
Bethe equation 
  (\re{be})  gives  
three distinct real  roots along the resonance line 
$\omega_f=\omega_a (\equiv \omega_{a1}=
\omega_{a2})  $, resulting to a { triplet} structure in 
the  VRS with  splittings in excitation   spectrum:
$E_{1}= 2.02 ,  3.02 , 4.02 $ for $\omega_f =3.02, \al =1$.
 Consequently,  the Rabi-oscillation   becomes  involved 
(see Fig.1a) with  three
transition frequencies. For small detuning $\de=\omega_f-\omega_a  $
the roots remain real, while    
at  $\de =\mp 0.30$ 
two of  them coincides, collapsing the Rabi-splitting to the usual doublet.
The excitation spectrum correspondingly reduces  to 
$E_{1}=  2.05,  3.81 \ (\mbox{degenerate})$  at sub-detuning and to 
$E_{1}=  2.22 \ (\mbox{degenerate}) , 3.99$
 at super-detuning points, reducing 
the  Rabi-oscillations 
 to   single frequency mode (Fig.1b). Beyond these detuning points   
  two of the roots  become
 complex conjugates, leading to an irregular 
  Rabi-oscillation
%, due to   decaying and enhancing solutions
 (Fig.1c).

\input{BSfig3r.tex}

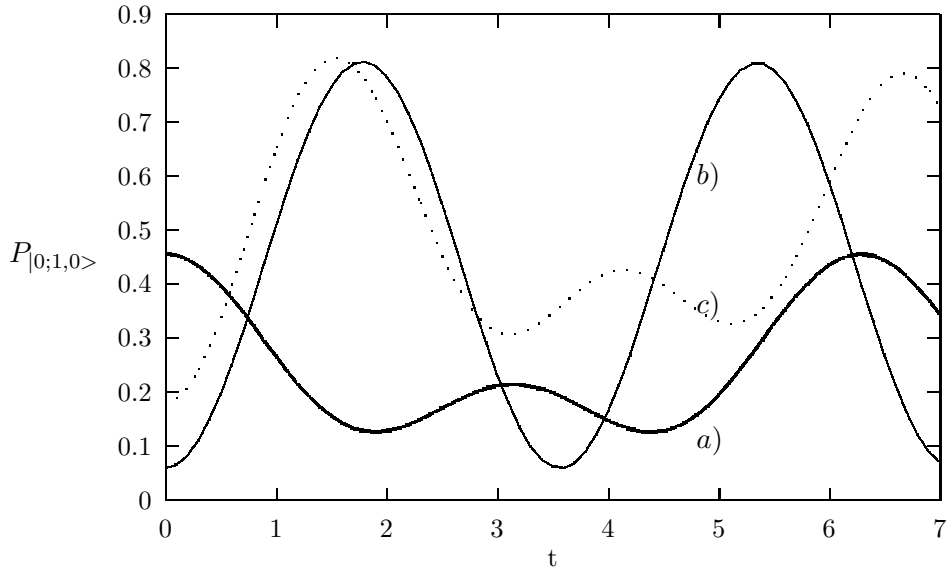
{\bf FIG. 1.} {\small Time dependence of transition
  probability,
 showing  Rabi oscillation
 at different detuning points: at a)   
   resonance,
% point involving three different frequencies,
b)   degenerate detuning 
%with  roots involving single frequency,
c)  further detuning with complex conjugate roots.}

Higher excitations for this model as well as its $N_a$-atom
extension  (\re{nbsh})
 can be solved exactly 
following the standard   BA formalism presented above. 
We explore now some subtle points and apparent  controversies
 regarding the BA solution of 
 BS models, which also have relevance for other integrable
models.
A common  belief, though  proved  only for specific  
 models \c{ndBA},
is that the degeneracy condition (i.e  $\la_a = \la_b$) 
for  the Bethe states:
 ${\al(\la_1) \ov \bet(\la_1)}=\pm 1$  
 can not be  solved apparently  for 
any integrable  model.
% and hence  Bethe states must  be  nondegenerate.
  We  however find that for our multi-atom 
BS model   at the resonance point $\de 
=\omega_f-\omega_a
=0$,
the degeneracy condition, which is equivalent to 
 ${\al(\la_1) = \bet(\la_1)}=0$ is indeed fulfilled,  
  yielding a  nontrivial 
solution $\la_1=\ha (- \omega_f \pm 1)$, which  
  recovers as well  the known  
 spectrum for the standard BS model: 
 $E_M=2M ( \omega_f-\la_1)=M E_1,$ at resonance \c{bs}.

Another apparent controversy regarding 
 multi-atom BS models, which  is also generic for many other models but
   not  emphasized properly in the literature, arises due to the fact
that, the dimension of the underlying Hilbert space for these models with
$N_a$ number of two-level atoms has a upper bound  $2^{N_a}$, for fixed
pseudoparticle number.
 For the standard BS model it is just $2$. Therefore
  the complexity  of the  problem
can not increase further  for  higher excitations with 
 $M > N_a$.
 For example, in the BS model
 by   diagonalizing  the Hamiltonian  directly  one can easily
get the exact energy spectrum 
for arbitrary excitation \c{bs} .
However,  when we  try to solve the same problem through Bethe ansatz
   the solution  must become
increasingly  complicated for higher excitations $|M>_B$ , since
 one has to find    all $M $  
 Bethe roots $\{\la _a\}, a=1,2,\ldots , M$ as solutions  to 
 general Bethe equations
({\re{be}}),
 which is impossible analytically! 
We  resolve this problem in an intriguing way by
  observing that Bethe state $|M>_B$ and the energy eigenvalues 
  depend in fact not on $M$ number of  
roots  $\la _a$ individually, but only on some
  symmetric  combinations of them and  moreover, the number of these relevant
variables does not exceed the dimension of the Hilbert space.
For the BS model for example, we find them to be  only two: $X_M,Y_M$ and
for deriving them  explicitly we
introduce  an equivalent set of symmetric Bethe roots through
 symmetric combinations of the original $M$ roots
\c{future}:
$s_1=\sum_a \la _a, \ s_2=\sum_{ab}=\la _a\la _b, \  \ldots ,
s_M=\prod_a \la_a$.
Combining suitably  BAE
({\re{be}}) and (\re{Lambda}) for the BS model,
 expressed through   symmetric roots
we arrive at the equations
\be
E_M X_M=\Delta_+ X_M+M Y_M, \ \ \
E_M Y_M=\Delta_- Y_M+M X_M, \ll{Em} \ee
where $\Delta_\pm =
M \omega_f\pm\ha \de$. On the other hand, expressing the Bethe
states  through two basic states of the model
 we find $|M>_B=X_M|M,->+Y_M|M-1,+>$, i.e.   dependant again
 on the above two relevant variables only.
 It is easy to see that the action of the
BS Hamiltonian on this eigenstate also  reproduces the same relation
(\re{Em}), we have derived  from the BAE.
 Fixing $ \omega_f=1$ for simplicity,
we find   these two variables  explicitly    through
 symmetric Bethe roots in the form
  $X_M=M s_M$ and $ Y_M=-Ms_M+ \ha \de(s_{M-1}+ \ldots +s_1 +1)$.
From the first of the relations (\re{Em}) we get
 the energy
spectrum  as  $E_M= \Delta_++M \kappa_M,$ where $\kappa_M={Y_M \ov X_M}$
 and using both these relations
derive the simple  equation $M\kappa_M ^2
 +{\de }\kappa_M -M=0$. This quadratic equation  is  solved easily
to yield  $E_M=M  \omega_f \pm (\de ^2 +M^2)^{\ha} $,
 recovering  the  known spectrum
of the BS model in the general $\de \neq 0$ case.
Thus through BA we get the explicit result for all higher excitations, also analytically, resolving the
raised controversy.
 Similar arguments
must hold in  the corresponding problem for other models.

Thus we have proposed through general Yang-Baxter algebra
 a series of new  integrable multi-atom
 matter-radiation models including q-deformed models and
  solved  them exactly
 through  Bethe ansatz in a unified way. Integrable trapped ion (TI)
and q-deformed TI models   introduced here are new, while  q-deformed
  Jaynes-Cummings (JC) and Buck-Sukumar (BS)    models
 are  multi-atom as well as integrable   extensions of
earlier models \c{qbs,qjc}. The
  proposed JC and BS models are nontrivial
generalizations of well known models \c{jcbethe,bsbethe},
 with the inclusion of inter-atomic interactions and
 nondegenerate atomic frequencies.
We find that,  contrary to the popular belief,
 the degenerate Bethe states do exist
 in the  multi-atom
BS models   at the resonance point.
 Multi-radiation modes can  be included   easily in such
  models preserving integrability.
Identifying the models in real systems
and  experimental verification  of
  the related results presented here, especially in many-atom microlasers
\c{n03},
 would be an important problem.
%------------------------------------

 \end{document}